
%
%
%
%
%

\magnification=1200
\baselineskip=13  truept
\font\norm=cmr10  scaled\magstep2

\vsize=23.8 true cm
\hsize=15. true cm
\hoffset 5. truemm
\topinsert
\noindent
December 1992 \hfill  Catania University preprint n. 92-03
\endinsert
\vskip 5 mm
\hrule
\vskip 2mm \noindent
\hrule
\vskip 20mm
\centerline{\bf  Cross section fluctuations and chaoticity}
\centerline{\bf  in heavy--ion dynamics }
\vskip 15mm
\centerline{   A.  Rapisarda }
\vskip 10mm
\centerline{\it INFN Sez. Catania, Corso Italia 57}
\centerline{\it I-95129 Catania, Italy}
\vskip 35mm
\hrule
\vskip 10mm
\centerline {\it Talk presented at the First
joint Italian--Japanese meeting }
\vskip 3mm
\centerline {\it "Perspectives in Heavy--Ion Physics" }
\vskip 3mm
\centerline {\it LNS Catania, Italy,
September 29 -October 2, 1992 }
\vskip 1 truecm
\centerline {\it (to be published in the
Conference Proceedings series of
the Italian Physical Society )}
\vfill
\eject
\topinsert
\vskip 1. truecm
\endinsert\noindent
\centerline {\bf CROSS SECTION FLUCTUATIONS AND CHAOTICITY }
\vskip 0.2 truecm
\centerline {\bf IN HEAVY--ION DYNAMICS }
\vskip 1.1 truecm
\noindent
\centerline{{ A.   Rapisarda}}
\vskip 1.6 truecm
\noindent
\centerline{ INFN Sez. Catania, Corso
Italia 57, 95129 Catania, Italy}
\vskip 2 truecm
\noindent
\centerline {\bf ABSTRACT }
\vskip .4 truecm
\noindent
{\it Cross section fluctuations in
nuclear scattering are briefly
reviewed in order to show the main important
features. Then chaotic scattering is
introduced by means of a very
simple model. It is shown that chaoticity produces
the same kind of irregular fluctuations observed
in light heavy--ion collisions. The
transition from order to chaos
allows a new general framework for a
deeper understanding of
reaction mechanisms.}
\vskip 10 truemm
\noindent
{\bf 1. INTRODUCTION}
\vskip 4 truemm
\par
In this presentation we will discuss
recent investigations about chaos
 in nuclear scattering. Though
chaos is not a new concept in
nuclear physics, only recently one is
completely realizing its important consequences [1].
The problem, expecially the quantal  aspect of it,
is rather elusive and delicate, however
 it shows very general features
which are of extreme interest for several fields [2-4].
Since most of the material
here presented has recently been
published in several papers [5-9] ,
we will not go into the
technicalities, which can
be found in the quoted references, trying on
the other hand to explain in a
simple and schematic language the reason
 of  chaoticity  onset, the meaning of it  and the
experimental implications according to
the present understanding.
This framework is still too young to draw
general conclusions, however it opens
new fascinating horizons,
connecting   nuclear physics with the
novel interdisciplinary research of
nonlinear dynamics and the evolution of complex systems.
\par The paper is organized as follows.
The main experimental results in the
scattering of light nuclei are
 briefly  summarized in section 2. The concept
of deterministic chaos
 is discussed in  section 3. Here
the classical and the quantal dynamics
of a simple  nuclear scattering problem
exhibiting chaoticity
are illustrated. Section 4 deals with a general discussion
on the implications of the transition from
order to chaos. A summary of
the main important results is done in section 5.
\vskip 16 truemm
\noindent
{\bf 2. EXPERIMENTAL BACKGROUND }
\vskip 4 truemm
\par
 Cross section fluctuations have been observed since the
 60s, when nucleon-nucleus reactions
started to be intensively studied
[10]. Predicted by Ericson [11,12], fluctuations
 in compound nucleus cross
sections  were detected at
excitation energies above the neutron
 evaporation  barrier. That is in the energy region
of strongly overlapping resonances, where the level spacing
D is very small in comparison
with the level width $\Gamma$, $\Gamma /D
\gg  1$. Fluctuations are generated
by the random action of the very
many intermediate levels which connect
the entrance and the exit channels. According
to Ericson's theory,
autocorrelation functions of experimental data
have a Lorentzian shape whose
width $\Gamma$, the coherence length,
gives the energy range within
which the intermediate levels are excited
coherently. Therefore $\Gamma$ represents the average
level width of the intermediate compound nucleus and
gives information on the average lifetime of the compound
nucleus $\tau=\hbar/ \Gamma$ and on the level density.
Fluctuations have a statistical nature, but are experimentally
reproducible. This
view is nicely confirmed by a vast literature [13].
However, experiments with heavier
projectiles - performed almost at the same
 time -  revealed  excitation function fluctuations with
different features. The first system to be studied
was $^{12}C + {^{12}C}$ [14]. In this case
fluctuations started around the Coulomb barrier presenting
structures with widths of different sizes.
In general these structures, which
were present in several reaction
channels,  became broader as the incident energy
increased. The coherence lengths extracted
from these experiments were larger than those
previously found in nucleon-nucleus scattering
- 100-300 KeV against
10-50 KeV-   and correlation analyses
showed a nonstatistical origin.
Similar characteristics were
observed for $^{12}C + {^{16}}O$ [15], and
$^{16}O + {^{16}O}$ [16] among several other systems [17].
Due to the peripheral kind
 of these reactions and the unusual strongly
attractive nucleus-nucleus potential at large
distances, it was postulated that these oscillating
structures should have a molecular-like nature
substantially  different from
that of the average compound nucleus.
Actually, some evidence of
this molecular origin has been found for
$^{12}C + {^{12}C}$  and $^{12}C + {^{16}O}$. At variance,
in other cases, for example $^{16}O + {^{16}O}$,
the situation remains more ambiguous [17].
\par Going to heavier
systems, a more complex behaviour was detected.
In correspondence of
excitation function fluctuations, anomalous large and highly
oscillating angular distributions
were observed. Typical examples of this behaviour
are the systems $^{16}O + {^{28}Si}$ [18] and $^{12}C +
{^{28}Si},{^{32}S}$ [19],
where these features were first measured.
Again a dinuclear molecule
was thought to be the origin, but the
mechanism soon appeared much more complicated:
systems leading to the same dinuclear composite showed
different structures;
 it was not always possible to understand
the angular distributions
in terms of only one single wave, on the contrary several
angular momenta around the grazing were involved [20].
The  phenomenon has been intensively studied
and, as in the case
of Ericson's fluctuations,
a  vast literature can be found on the subject.
Fundamental review papers, both
on the many experiments performed and  the  theoretical
models proposed  to explain heavy--ion resonances,
are those of Erb and Bromley [17]
and Braun-Munzinger and Barrette [21].
They say clearly that
fluctuating phenomena in light
systems seem to have a common nature:
there are only quantitative, but not qualitative
differences  from system to system.
However, notwithstanding the
great effort spent during these
years, there is not yet
a quantitative theoretical understanding
of this behaviour: all the
advanced models fail - partly
or completely - in reproducing the
large set of existing data.
The only model-independent consideration which
comes out naturally from
the experimental analysis is the unexpected
presence of a very weak surface absorption.
In other words, a  relatively small
number of channels is involved.
\par Though the interest in these
mysterious phenomena diminished
in the 80s, some groups have continued
 the experimental research. Thus
fluctuations were recently found in the
elastic and inelastic cross sections of heavier nuclei like
$^{28}Si + {^{28}Si}$ [22],
$^{24}Mg + {^{24}Mg}$ [23] and $^{24}Mg + {^{28}Si}$ [24].
At the same time excitation function
fluctuations were observed also in
deep inelastic collisions  of
several systems like $^{19}F + {^{89}Y}$
 [25], $^{28}Si + {^{64}Ni}$, $^{28}Si + {^{48}Ti}$ [26].
Again, differences from Ericson's theory were found,
mainly because correlations between
several channels and a
clear angular dependence were evidenced.
In ref. [27] cross section
fluctuations were measured for several
windows of energy loss, establishing
this way a connection between
oscillating  phenomena in elastic and damped  reactions.
Further aspects on the current
experimental studies of excitation
function fluctuations in light
nuclei can be found in the contributions
by  G. Pappalardo and D. Vinciguerra to this conference.
\par This is very briefly the puzzling state of the art
of the experimental fluctuations in light nuclei collisions,
which is still waiting  for a
sound theoretical comprehension.
 In what follows the concept of chaotic scattering will
be introduced and one will see that this new
perspective is able to justify
the experimental phenomenology presented above.
\vskip 6 truemm
\noindent
{\bf 3. CHAOTIC SCATTERING }
\vskip 4 truemm
\par
The intensive studies on nonlinear
dynamical systems have demonstrated that
simple deterministic laws can exhibit a
complex and unpredictable evolution
for infinitesimal variations of the
initial conditions. This irregular
behaviour, which was firs
discovered by Poincar\'e at the end of
the last century [28],  but has been thoroughly
investigated only since the 70s, goes under the name of
{\it deterministic chaos} [29].
In general, considering classical  hamiltonian systems,
one should expect chaoticity when
the system is {\it non-integrable}, i.e.
the number of degrees of freedom is greater than
that of the constants of motion. In
other words  chaos shows up
if a symmetry breaking occurs
and a strong coupling between the
equations of motion exists.
Let us consider for example the following
 hamiltonian describing a generic two-body problem
whose interaction has a central and a non-central term
$$
       H(r,p,\theta,I)=
{p^2\over 2m} + {I^2\over 2\Im} + V(r) +
\alpha U(r,\theta) \, . \eqno(1)
$$
Here p is the momentum,
m the reduced mass, I the angular momentum,
 $\Im$  the moment of inertia,
 $V(r)$ the central potential and
$\alpha$ the strength of the
non-central potential $U(r,\theta)$.
If $\alpha=0$
the system is {\it integrable}: we have 2 degrees of
freedom, $r$ and $\theta$,
and 2  conserved quantities, the energy $E$ and the
angular momentum $I$. It
is possible to solve exactly our problem by decoupling the
classical hamiltonian  equations. But if $\alpha \neq 0$ then
the rotational symmetry is destroyed,  the only conserved
quantity is the energy  $E$ and it is not possible to find
an exact solution. The system can show an irregular behaviour
whose degree of chaoticity depends on
the strength $\alpha$. The transition from a
ordered dynamics to a completely chaotic one is gradually
regulated by the coupling strength.
In general
the two extreme cases, {\it order} ($\alpha=0$)
  and {\it hard chaos}
($|\alpha| \gg 0$ ), are rather rare. On the other hand,
very often one has to deal with a mixed behaviour, called
{\it soft chaos}, where  $\alpha$  is not so sufficiently small
that perturbation theory can be applied [2,30].
\topinsert
\vskip 8.5 truecm
\noindent
\settabs 2 \columns
\+  Figure  1. Final scattering angle \quad $\phi_f$ &
 Figure 2. Deflection function for a fixed \cr
\+ as a function of the initial orientation
   & orientation angle $\theta_i$ (a). \quad In (b) and (c) \cr
\+ angle \quad $\theta_i$ for three different incident
   &  two blow-ups are shown in order to show \cr
\+ energies.
   & the fractal structure. See text.    \cr
\vskip 5 truemm
\endinsert
\noindent
In quantum mechanics, due to the limitations imposed by the
uncertainty principle, such an irregular behaviour should not be
present. This is one of the most debated topics in
the current literature [2-4].
According to the present understanding of the phenomenon,
the transition from order to chaos occurs  also in quantum
mechanics, but it exhibits  smoother features substantially
different from the fractal structure of  classical chaos.
In general one calls {\it quantum chaos} the
quantal analogue of
those systems which are classically chaotic.
\par
Coming back to the classical case and to eq. (1), we can study
a bounded dynamics or a scattering problem.  Strictly speaking,
the scattering problem is integrable at large distances
where the potential goes to zero. However, due to the fact
that one probes an interaction region
which is chaotic, the final observables
(final scattering angle, final angular momentum, etc.)
present wild oscillations at all scales and a mixture of
regular and irregular islands as
a function of the initial conditions.
This phenomenon
is called {\it chaotic scattering} [2,31-34].
\par
In ref.[5,6]   it has been proved that, when considering
a deformed nucleus impinging on a spherical one,
chaotic scattering
occurs  if the nuclei are not very heavy
and one is not very far from the grazing
condition.  The hamiltonian in this case can be written
as
$$
{\cal H} \,  =\,  {p^2\over{2 m}} + {I^2 \over {2\Im}} +
{(L-I)^2\over{2 m r^2}} + V(r,\theta)      \, ,
\eqno (2)
$$
\noindent
where $ L$ and $I$ are respectively the total angular momentum
and the spin of the rotor.  The potential $V(r,\theta)$ is the
sum of the Coulomb and the nuclear interaction.
In particular, as a typical example, the
system $^{28}Si+ {^{24}Mg}$ was considered
(see refs. [5,6] for further details).
Solving numerically the equations of motion, one gets for the
final observables the fluctuations
shown in figs. 1 and 2.
In fig.1, for the  $^{28}Si+ {^{24}Mg}$ reaction, the
final scattering angle
$\phi_f$  is displayed as a function of the initial
orientation $\theta_i$. Three different incident energies
are considered. Each plot contains 1800
points and each point is the result of the numerical
integration for a particular  initial orientation.
This irregular behaviour occurs at all scales as fig.2 shows
for the case of the  deflection function
(see also figs. 3,4 of ref.[6]).
Here the final scattering angle $\phi_f$ is reported as
a function of the total angular momentum $L$,  for a fixed
incident energy and initial orientation angle $\theta_i$. The
value of $L$ corresponds to the initial orbital
angular momentum $\ell_i$ being $I_i$=0. In the same
figure two blow-ups of smaller regions are shown
in order to  display the fractal structure,
characteristic signature of chaoticity.
\par
These impressing chaotic features are not due to the
peculiarities of our model and have been recently
found also in a different nuclear scattering  problem
considering vibrations  as the only degrees of freedom [35].
In general, such fractal fluctuations are generated by
trapped trajectories which remain in the
internal pocket for a very long time. Since the interaction
zone is chaotic,
it is not possible to predict this
trapping time (or delay time) which depends in a
very sensitive way on the initial conditions. This is
illustrated in the schematic pictorial
view of fig.2. Here the oscillations of the potential, due to the
coupling of the relative motion to the rotor spin, and the
chaotic evolution
of one trajectory are shown together with a generic excitation
and dexcitation of the internal degrees of freedom.
In this scheme the absorption has  been neglected, but from
 the experimental overview of the previous
section, one knows that this is
not a bad approximation for light heavy--ion systems. The role
of the absorbtion will be discussed
later in connection with the quantal  dynamics.
\par Concluding the review of the classical dynamics,
one can claim that irregular scattering is a general
phenomenon which coexists
with a regular one when - and this occurs very often - the
problem is non-integrable.
\topinsert
\noindent
\settabs 2 \columns
\+ & Figure  3. Pictorial view of chaotic scat-        \cr
\+ & tering. Due to the strong coupling, the  \cr
\+ & interaction $\,$ region
shows $\,$ large $\,$oscilla- \cr
\+ & tions  for infinitesimal $\,$ variations of the  \cr
\+ & initial $\,$ conditions. $\,$ Therefore the inci-\cr
\+ & dent $\,$ nucleus is $\,$ trapped and $\,$ excited, \cr
\+ & until $\,$ it $\,$ finally
$\,$ succeeds $\,$ in escaping  \cr
\+ & again. Of course, $\,$ classically $\,$ the inter- \cr
\+ & nal degrees of freedom are not discrete. \cr
\vskip 9 truecm\noindent
 Figure 4. Quantal elastic transition probability as a function
of incident energy.  The calculations were
performed by means of a  2-d coupled-channels approach,
which is the quantal analogue of the classical chaotic
scattering problem.  The energy step
used in the calculations is   20 KeV. See text.
\vskip 5 truemm
\endinsert
\noindent
One should not forget that
nuclei are quantum objects and
so the question is now if this new perspectives can
influence the quantum description.
This is expecially important when the semiclassical approximation
 can be applied. In the following the quantal analogue
of the above discussed classical model
will be reviewed.
Several recent studies on dynamical systems have  demonstrated
that the classical transition from order to chaos
has a quantal analogue [2-4, 31,32,36]. We will see that this
is the case also in our model.
\topinsert
\vskip 8 truecm
\noindent
\settabs 2 \columns
\+  Figure  5. Excitation functions calcu-
  &   Figure 6. Autocorrelation functions of the \cr
\+ lated for  the system $^{28}Si + {^{24}Mg}$ by
   &  fluctuations displayed in fig.5 (full curves). \cr
\+ means of the code  FRESCO [38].
 & The \quad dashed curves are \quad Lorentzian fits \cr
\+ The c.m.   angle is 178$^o$. See text.
  & whose widths are  reported in the figure.  \cr
\vskip 5 truemm
\endinsert
\noindent
The quantal approach which corresponds to the above discussed
 classical scattering
is the  2-dimensional coupled-channels model studied
in refs.[6-9]. These investigations
have demonstrated that the quantal analogue of chaotic
scattering is the appearance of
sharp irregular fluctuations in the transition probabilities
as a function of energy.
The irregular structures are caused by the presence
of an asymmetric pocket and a strong coupling term between
the different channels [9].  It has been checked that
these quantal irregularities are present
in the same energy and angular region where classical chaos
shows up.  More precisely fluctuations
start at the potential barrier and their widths become broader
and broader until they disappear completely by increasing energy
and/or  angular momentum.
An example of these oscillations, for the same system
investigated in
the classical case, is displayed in fig.4. In the figure only the
elastic channel is displayed. The total
angular momentum is $L$=15 $\hbar$, while the initial
 spin of the rotor is $I$=0.
The energy step used was 20 KeV, no structures
below this energy step were found [6-9].
Similar features have been obtained taking into
account vibrations  in a nuclear reaction [37].
Actually, various chaotic  scattering problems [31,32,36]
show an identical behaviour.
\par
A more quantitative analysis of the
fluctuations illustrated in fig.4 was done by investigating
the autocorrelation functions,
both in the classical and in the quantal case [8,9]. The
coherence lengths
extracted in the two cases were comparable within a
factor of two and of the order of 100-250 KeV.
 Therefore as in ref.[31,32],
a firm correspondence exists between the classical and
the quantal dynamics,
notwithstanding the semiclassical theory is not valid a
priori in the problem into consideration.
Unfortunately, the small number of resonances has not
 permitted the study of their distribution as in
[31] where a GOE law was evidenced.
\topinsert
\vskip 11 truecm
\noindent
Figure  7. Elastic angular distributions,
with $86^o < \theta_{cm} < 178^o$,
as a function of incident energy
calculated by means of the code FRESCO [38], see text.
\vskip 5 truemm
\endinsert
\noindent
The irregularities in the transition probabilities are
reflected drastically in the
excitation functions and the angular distributions as
a function of energy.
This result has been checked in refs.[8,9],
where it was shown how transition probabilities summed
over different incident waves continued to
display irregular fluctuations. More realistic 3-dimensional
calculations were performed [9] by means of the code FRESCO [38].
They confirmed the presence of irregular fluctuations above the
Coulomb barrier, both in the excitation functions and in the
angular distributions. Fig.5 displays  these cross section
fluctuations as a function of energy for the
channels considered, i.e. only the states
$0^+,2^+,4^+$ in $^{24}Mg$.
Fig. 6 shows the correspondent autocorrelation functions
(full curves) with
the Lorentzian fits (dashed curves) and the extracted
coherence lengths.
A preliminary analysis of the cross sections reported in fig.5
has revealed some correlation
among the considered channels. On the other hand, as fig.6 shows,
there are clear deviations from
the Lorentzian shape
predicted by Ericson's statistical theory [9].
These are important points  which
will be discussed in the next section.
In fig.7 the angular distributions are shown
as a function of the incident energy.
This plot illustrates nicely how,
analogously to the experimental data,
theoretical calculations exhibit an
irregular behaviour which starts
at the Coulomb barrier and is more evident at
backward angles, where the absolute
value of cross section is smaller [9].
The cross sections of figs.5 and 7 were calculated using  a small
imaginary part of the potential, according
to ref.[39] and to the experimental evidence [17,21]. In ref.
[7] it was shown that the strength of
the absorption is a crucial point, since
it can smooth and even wash out completely the
fluctuations.
\par The 3-d calculations give therefore a satisfactory
justification of the
experimental anomalies discussed in section 2. At the
same time, on the other hand, they
connect the experimental data to
chaotic scattering, being the code FRESCO a more
refined version of the
2-d quantal model whose classical analogue is chaotic.
\par
Concluding this section, it is important
to note that preliminary calculations
for the systems $^{12}C + {^{24}Mg}$ and $^{16}O + {^{28}Si}$
have revealed the same features [40].
\vskip 6 truemm
\noindent
{\bf 4. DISCUSSION}
\vskip 4 truemm
\par
The phenomenology which comes out of chaotic
scattering calculations is therefore identical to that
observed experimentally for light nuclear systems.
These theoretical investigations
indicate strongly that the experimental
fluctuations have a chaotic origin. At the same time
 they demonstrate the smoother but impressing sensitiveness
of quantum chaos to small changes of the input parameters.
In this sense chaos maintain a clear
identification also in quantum mechanics.
In refs. [27,31] the chaotic  origin of the scattering between
light nuclei was already claimed. In this presentation
and in refs.[5-9] this hypothesis has been
 investigated in detail and a sound support has been
given to it. It is true, however,  that
a systematic investigation and a more
complete quantitative analysis is needed.
It should be stressed that
a quantitative study has a meaning only
within a statistical analysis. In general, in fact,
 it has no sense
trying to reproduce the single fluctuation produced by the
interplay of several resonances.
We are in a regime where
minimal - even if not infinitesimal - changes of
the input parameters have a strong influence in the
final results.
Our knowledge  of nuclear physics is not so good to trust
the parameters
used to that accuracy and several
parametrizations are able to fit the same structures [39].
\par
If it is true that this theoretical  analysis is only at
the beginning,
more refined experiments should be also performed.
The nowadays available $4\pi$ detectors and more sophisticated
experimental techniques could give a crucial contribution
in order to draw definitive conclusions.
 In particular irregular fluctuations should be detected
also in the gamma multiplicities
as preliminary chaotic scattering calculations  indicate [40].
\topinsert
\hrule
\vskip 0.5 truecm
\settabs 3 \columns
{\norm
\+ { \quad ORDER} & elastic scattering &   optical model  \cr
\+   &  &  \cr
\+ $\quad\quad\quad \Big\Downarrow$  & $\quad\quad \quad
\Big\Downarrow$
     & $\quad\quad \quad \Big\Downarrow$  \cr
\+   & non-equilibrated  &   \quad \quad TDHF \cr
\+ {\norm MIXED REGIME} & and  rotating
& coupled-channels    \cr
\+  & dinuclear system  &  transport theories   \cr
\+ $\quad\quad\quad\Big\Downarrow$
& $\quad\quad \quad \Big\Downarrow$
 & $\quad\quad \quad \Big\Downarrow$  \cr
\+  &  &   \cr
\+ { \quad CHAOS }  &  excited compound
& Random Matrix Theory  \cr
\+  & nucleus  & Ericson's fluctuations  \cr}
\vskip 0.5 truecm
\hrule
\vskip 1. truecm
\noindent
Table 1. Schematic table illustrating the
 different stages of the transition from order to chaos
($1^{st}$ column) in correspondence of
the physical picture ($2^{nd}$ column) and the standard
models use in nuclear reaction theory ($3^{rd}$ column).
\vskip 5 truemm
\endinsert
\noindent
\topinsert
\hrule
\vskip 0.5 truecm
\settabs 3 \columns
{\norm
\+ {  Direct reactions }
   & Complex reactions
   & Compound reactions    \cr
\+       &    &                        \cr
\+       &    &                        \cr
\+       &    &                        \cr
\+       &    &                        \cr
\+       &    &                        \cr
\+       &    &                        \cr
\+ $\quad\quad\quad \Big\Updownarrow$
    & $\quad\quad \quad \Big\Updownarrow$
   & $\quad\quad \quad \Big\Updownarrow$  \cr
\+       &    &                        \cr
\+       &    &                        \cr
\+  \quad {$\tau\sim10^{-22} \, s.$} &
\quad {$\tau\sim10^{-21} \, s.$} &
\quad {$\tau\sim10^{-20} \, s.$ }\cr
\+       &    &                        \cr
\+  \quad {$\Gamma\sim 2-4 \, MeV$}     &
\quad {$\Gamma\sim 200-400 \,  KeV$ }      &
\quad {$\Gamma\sim 10-50 \, KeV$ } \cr}
\vskip 0.5 truecm
\hrule
\vskip 1. truecm
\noindent
Figure  8. Pictorial view of the three kinds of reactions
which contribute to the cross sections in the mixed regime
({\it soft chaos}).
The characteristic coherence widths and reaction
times are also displayed.
\vskip 5 truemm
\endinsert
\noindent
Therefore chaotic scattering outlines
a new framework in which old puzzles find
their natural justifications and a
deeper insight of nuclear reaction mechanisms can
be gained.
In particular two important considerations should be underlined.
The first is that  a few degrees
of freedom are able to produce a very complicated dynamics. This
is a very general warning and should be always borne
in mind, expecially when the semiclassical
approximation is used. The second is that
both {\it order} and {\it chaos} are two extreme regimes.
We know very well how to describe these two cases. In
the former the optical model works pretty well,
while in the latter statistical theories,
like Random Matrix Theory or the  Ericson one,
can be successfully used [1,31,41].
In reality the most general situation
is very often in between [2]. This mixed regime -
 {\it soft chaos} -
is , however, the most difficult case to treat:
fluctuations are not completely statistical and
dynamical correlations exist. The problems
discussed in section 2 and 3
seem to belong just to this regime: both
the experimental data and the dynamical  model
presented show deviations from Ericson's theory.
That  is why several models in the past,
 like for example time-dependent Hartree-Fock,
coupled-channels approaches and transport theories,
 have had only a partial
success: in general only average quantities
have been correctly described.
The transition from order to chaos  allows
to reinterpret standard models offering, at
the same time, a deeper dynamical insight.
Table 1 summarizes this schematic discussion, illustrating
the different physical situations  ($2^{nd}$ column)
in correspondence of the
chaoticity regime ($1^{st}$ column) and
of some well-known theoretical approaches ($3^{rd}$ column).
\par
Coming back to Ericson's fluctuations and
Random Matrix Theory, in
our view, they apply only in the complete
chaotic regime. In this sense the excited
compound nucleus reactions are nice examples of chaotic
behaviour [41,42]. However, from the above discussion,
one should expect deviations in the case of soft chaos.
Actually, since the first papers [12], Ericson
himself warned that his formulation
did not take into account possible coherence effects,
which could generate interference between {\it direct} and
{\it compound} reactions.
In general, is not always possible
to decompose the cross section into a direct term and
a compound one. This is what happens in the mixed regime,
where a third kind of reactions, we can call them {\it complex
reactions}, should be considered. In
the classical picture these are those who create the
islands of regularity  inside the chaotic regions. Such
collisions not only  can enlarge the coherence length,
having different reaction
times, but they can often produce small oscillations
in the autocorrelation function at energies
greater than the coherence length [25,26,40,43]. The
three different kind of reactions are illustrated
in a pictorial view
in fig.8, where also the characteristic
coherence lengths and reactions times are reported.
\par
This new scenario may seem too generic
and chaotic scattering a superficial rephrasing
of old theories or an easy
explanation for unsolved problems. In reality,
though a lot has still to be done in order
to be more precise and characterize the
different stages with an increasing
degree of chaoticity (or complexity),
the chaotic framework is the
new emerging interdisciplinary paradigm of natural sciences.
It is a novel powerful language whose limits are
still unknown, which can refresh
nuclear physics and be successful
where the old way of thinking failed [44,45].
\vskip 6 truemm
\noindent
{\bf 5. SUMMARY}
\vskip 4 truemm
\par
It has been shown that chaotic scattering represents
a real possibility in collisions between light
nuclei and that it can explain
the irregular fluctuations observed experimentally.
This is an important result both for nuclear physics
and for more fundamental questions
like the existence and the features of
quantum chaos. These investigations
allow to reinterpret standard approaches - although
for the moment only in a generic way -
in the new framework of the
transition from order to chaos.  The study of
heavy--ion scattering is particularly interesting due to its
privileged position  between the classical and
the quantum world.
\vskip 5truemm
This contribution is part of a fruitful collaboration with
 M. Baldo and
E.G. Lanza. The author thanks M. Papa, G. Pappalardo
and D. Vinciguerra for
several stimulating discussions.
\vfill\eject
{\bf  REFERENCES }
\vskip 4 truemm
\item{[1]} O. Bohigas and H.A.
Weidenm\"uller, Ann. Rev. Nucl. Part.
Sci. 38 (1988) 421 and references therein.
\par
\item{[2]} M.C. Gutzwiller, \quad
{\it Chaos in Classical and Quantum
Mechanics}, \quad  Springer-Verlag,
1990, and references therein.
\par
\item{[3]} R.V. Jensen, Nature 355 (1992) 311.
\par
\item{[4]} Proceedings of the International School on
{\it Quantum Chaos}, Varenna (Italy) 1991,
Eds. U. Smilansky and G. Casati, in press.
\par
\item{[5]} A. Rapisarda and M. Baldo,
Phys. Rev. Lett. { 66} (1991) 2581;
\par
\item{$\,$}  A. Rapisarda and M. Baldo,
Proceedings of the International Workshop on
{\it Intermediate Energy Reaction Processes
and Chaos}, Varenna (Italy),
29 september - 5 october 1991, in press.
\par
\item{[6]}  A. Rapisarda and M. Baldo,
Proceedings of the
{\it XXX Bormio International  Winter Meeting},
Bormio (Italy) 1992, Ed. I.Iori, Ric.
Scient. Educ. Perm. suppl. n.91, p.354.
\par
\item{[7]}  M. Baldo and A. Rapisarda,
Phys. Lett. B 279 (1992) 10.
\par
\item{[8]}  M. Baldo and A. Rapisarda,
Phys. Lett. B 284 (1992) 205.
\par
\item{[9]}  M. Baldo, E.G. Lanza and
A. Rapisarda , Nucl. Phys. A
545 (1992) 467c.
\par
\item{$\,$}  M. Baldo, E.G. Lanza and A. Rapisarda,
Proceedings of the International Workshop
{\it From Classical to Quantum Chaos
(1892-1992)}, july 21-24 1992,
SISSA Trieste (Italy), to be published.
\par
\item{[10]} L. Colli, U. Facchini,
I.Iori, M.G. Marcazzan, M. Milazzo,
and F. Tonolini, Phys. Lett. 1 (1962) 120;
\par
\item{$\,$}  Y. Cassagnou, Mlle I. Iori,
Mme C. Levi, T. Mayer-Kuckuk, M. Mermaz and
Mme Papineau, Phys. Lett. 6 (1963) 20;
\par
\item{$\,$} P. Von Brentano, J. Ernst,
O. H\"auser, T. Mayer-Kuckuc,
A. Richter and  W. Von Witsch Phys. Lett. 9 (1964) 48.
\par
\item{[11]} T. Ericson, Phys. Rev. Lett. {5} 30 (1960) 430;
\par
\item{$\,$} D. Brink and R. Stephen,
Phys. Lett. { 5} (1963) 77.
\par
\item{[12]} T. Ericson, Ann. Phys. 23 (1963) 390.
\par
\item{[13]} T. Ericson and T. Mayer-Kuckuk, Ann. Rev.
Nucl. Sci. 16 (1966) 183 and references therein.
\par
\item{[14]} D.A. Bromley  J.A. Kuenher and E.
Almqvist Phys. Rev. Lett. 4 (1960) 365.
\par
\item{[15]} M.L. Halbert, F.E. Durhnam and A. van der Woude,
Phys. Rev. 162 (1967) 899.
\par
\item{[16]} R.H. Siemssen  J.V. Maher  A. Weidinger and
D.A. Bromley Phys. Rev. Lett. 19 (1967) 369.
\par
\item{[17]} K. A. Erb  and D. A. Bromley,
{\it Treatise on  heavy--ion
 Science}, vol.3, Ed. by D. A. Bromley ,
Plenum Press (1984), New York.
\par
\item{[18]} P. Braun-Munzinger, G.M. Berkowitz, T.M.
Cormier, C.M. Jachcinski, J.W. Harris, J. Barrette and M.J.
Le Vine, Phys. Rev. Lett. 38 (1977) 944;
\par
\item{$\,$} J. Barrette, M.J. Le Vine,
 P. Braun-Munzinger, G.M. Berkowitz, M. Gai,
J.W. Harris,   C.M. Jachcinski, Phys. Rev.
Lett. 40 (1978) 445;
\par
\item{$\,$} P. Braun-Munzinger,
G.M. Berkowitz, M.Gai, T.M. Cormier, C.M. Jachcinski,
T.R. Renner, C.D. Uhlhorn, J. Barrette and M.J.
Le Vine, Phys. Rev. C 24  (1981) 1010.
\par
\item{[19]} R. Ost, M.R. Clover,
R.M. De Vries, B.R. Fulton, H.E. Gove,
and N.J. Rust Phys. Rev. C 19 (1979) 740.
\par
\item{[20]}  M.C. Mermaz, A. Greiner, B.T. Kim,  M.J. LeVine,
E. M\"uller, M. Ruscev, M. Petrascu,
M. Petrovici, and V. Simion, Phys. Rev. C 24 (1981) 1512.
\par
\item{$\,$}   M.C. Mermaz, E.R. Chovez-Lomeli,
J. Barrette, B. Berthier and A. Greiner,
Phys. Rev. C 29 (1984) 147;
\par
\item{[21]} P. Braun-Munzinger and
J. Barrette, Phys. Rep. 87 (1982) 209.
\par
\item{[22]}  R.R. Betts, B.B. Back and B.G.
Glagola, Phys. Rev. Lett. 47 (1981) 23.
\par
\item{[23]}  R.W. Zurm\"uhle, P. Kutt, R.R. Betts, S. Saini,
F. Haas and O. Hansen, Phys. Lett. B 129 (1983) 384.
\par
\item{[24]} A.H. Wuosmaa, S. Saini, P.H. Kutt, S.F. Pate,
R.W. Zurm\"uhle and R. R. Betts,
Phys. Rev. C { 36} (1987) 1011;
\par
\item{$\,$} A. Sarma and R. Singh, Zeit.
f. Phys. { 337} (1990) 23.
\par
\item{[25]} T. Suomij\"arvi, B. Berthier, R. Lucas,
M.C. Mermaz, J.P. Coffin, G. Guillaume,
B. Heusch, F. Jundt and F. Rami, Phys. Rev. C 36 (1987) 181.
\par
\item{[26]} G. Pappalardo, Nucl. Phys. A{ 488} (1988) 395c;
\par
\item{$\,$} G. Cardella, M. Papa, G. Pappalardo,
F. Rizzo, A. De Rosa,
G. Inglima and M.Sandoli, Z. Phys. A 332 (1989) 195;
\par
\item{$\,$} G. Cardella, M. Papa, F. Rizzo, G. Pappalardo,  A.
De Rosa,
E. Fioretto, G. Inglima, M. Romoli,  and M.Sandoli,
Proceedings of the
{\it XXX Bormio International  Winter Meeting},
Bormio (Italy) 1992, Ed. I.Iori,
Ric. Scient. Educ. Perm. suppl. n.91, p.140, and references
 therein.
\par
\item{[27]} A. Glaesner, W. D\"unweber, M. Bantel, W. Hering,
 D. Konnerth,
R. Ritzka, W. Trautmann, W. Trombik and W. Zipper, Nucl.
Phys. A   509 (1990) 331.
\par
\item{[28]} H. Poincar\'e, {\it Les Methodes Nouvelles de
la Mechanique Celeste}, Gauthiers-Villars, Paris, 1892.
\par
\item{[29]} See for example in the case of hamiltonian systems:
M. Tabor, {\it Chaos and Integrability in Nonlinear Dynamics},
J. Wiley, 1989.
\par
\item{$\,$} See for example in the case of dissipative systems:
P. Cvitanovic, {\it Universality in Chaos}, Adam Hilger, 1989.
\par
\item{[30]} I. C. Percival, Proceedings of the Royal
Society of London, Eds. M. V. Berry, I. C. Percival and
N. O. Weiss, Princeton University Press, 1987, p.131.
\par
\item{[31]} R. Bl\"umel and U. Smilansky,
Phys. Rev. Lett. { 60} (1988) 477;
\par
\item{$\,$} U. Smilansky, Proceedings of the International
School on {\it Chaos and Quantum Physics},
Les Houches 1989, Eds. M.J. Giannoni, A. Voros and
J. Zinn-Justin, Elsevier Science
Publishers B.V., 1991, and references therein.
\par
\item{[32]} P. Gaspard and S.A. Rice,
J. Chem. Phys. 90 (1989) 2225,2242, 2255.
\par
\item{[33]} C. Jung and H.J. Scholz, J. Phys. A 20 (1987)3607;
\par
\item{$\,$} C. Jung and T. T\'el, J. Phys. A 24 (1991) 2793;
\par
\item{$\,$} B. Eckhardt, Physica D 33 (1988) 89.
\par
\item{[34]} S. Bleher, C. Grebogi and E. Ott,
Physica D 46 (1990) 87.
\par
\item{[35]} C.H. Dasso, M. Gallardo and M.
Saraceno, Nucl. Phys. A 549 (1992) 265.
\item{[36]} K. Ohnami and Y. Mikami J. Phys. A 25  (1992) 4903.
\par
\item{[37]} C.H. Dasso,
M. Gallardo and M. Saraceno, in preparation.
\par
\item{[38]} I.J. Thompson, Comp. Phys. Reps. 7 (1988) 167.
\par
\item{[39]} G. Pollarolo and
R.A. Broglia, Nuov. Cim. 81 A 278 (1984).
\par
\item{$\,$} V.N. Bragin, G. Pollarolo and
A. Winther, Nucl. Phys. A { 456} 475 (1986).
\par
\item{[40]} M. Baldo, E.G. Lanza
and A. Rapisarda, in preparation.
\par
\item{[41]} H. Weidenm\"uller, Nucl. Phys. A 518 (1990)
1 and references therein.
\par
\item{[42]} Y. Alhassid and
D. Vretenar, Phys. Rev. C 46 (1992) 1334;
\par
\item{[43]} S. Yu. Kun, Phys. Lett. B 257 (1991) 247.
\par
\item{[44]} V.G. Zelevinsky, Proceedings of the
International Nuclear
Conference, July 26 - August 1 1992, Wiesbaden (Germany),
Nucl. Phys. A in press.
\par
\item{[45]} Proceedings of the International Conference on
{\it Dynamical Fluctuations and Correlations in
Nuclear Dynamics}, Nucl. Phys. A 545 (1992).
\vfill
\bye